\documentclass[manuscript]{aastex}

\slugcomment{To appear in Astrophysical Journal} 

\shorttitle{Dust in  clusters}
\shortauthors{Guti\'errez \& L\'opez-Corredoira}

\begin{document}

\title{Dust in  clusters: separating the contribution of galaxies and  intracluster media}

\author{C. M. Guti\'errez\altaffilmark{1,2} and M. L\'opez-Corredoira\altaffilmark{1,2}}
\affil{$^1$ Instituto de Astrof\'\i sica de Canarias, 38205 Tenerife, Spain; $^2$
University of La Laguna, 38206 Tenerife, Spain}

\email{cgc@iac.es}

\begin{abstract}
We have analized a sample of 327 clusters of galaxies spanning the range
0.06-0.70 in redshift. Strong
constraints on their mean intracluster emission by dust have been obtained using
maps and catalogs from the HERSCHEL HerMES project; within a radius of 5 arcmin
centered in each cluster, the 95\% C.L. limits obtained are 86.6, 48.2 and 30.9
mJy at the observed frequencies of 250, 350 and 500 $\mu$m.  From these restrictions, and assuming 
physical parameters typical of interstellar media in the Milky Way, we have obtained tight upper limits on the
visual extinction of background galaxies due to the intracluster media: $A_V(95
\%\, C.L.) \lesssim 10^{-3}$ mags. Strong constraints are also obtained for the mass of such
dust; for instance using the data at 350 $\mu$m we establish a 95\% upper limit
of $<10^9M_\odot$ within a circle with a radius of 5 arcmin centered in the
clusters. This corresponds to a fraction of the total mass of the clusters of
$9.5\times 10^{-6}$, and indicates a deficiency in the gas-to-dust ratio in the
intracluster media by about three orders of magnitude as regards the value found
in the Milky Way. Computing the total infrared luminosity of the clusters in
three ranges of redshift (0.05-0.24, 0.24-0.42 and 0.42-0.71) and two ranges of
mass ($<10^{14}$ and $>10^{14}M_\odot$) respectively, a strong evolution of
luminosity in redshift ($L\sim z^{1.5}$) for both ranges of masses is found.
The results indicate a strong declining in star formation rate with time in the
last $\sim 6$ Gyr.

\end{abstract}


\section{Introduction}

Dust 
attenuates the light of background objects, it is  a key ingredient in the process of
star formation and plays relevant roles in mechanisms of cooling and spreading out of metals
in galaxies. 
The presence of dust in the interstellar media of the Milky Way has been known since
nearly a century ago, and its emission mapped with relatively good sensitivity and angular
detail by the satellites IRAS  \citep{whe93} and COBE DIRBE \citep{ben92}. From those
maps, \citet{sch98}  built a map of optical extinction in  the Milky Way that has been
widely used. Although dust is routinely detected in extragalactic environments, the
existence of dust  in intracluster media (ICM) is still a matter under discussion.  There are
several processes that expel dust out of the galaxies, and  some of them which involve
the presence of at least two galaxies, like ram pressure stripping or mergers, are probably
more common within clusters of galaxies. However, it is unclear if such continuous or
episodic injection of dust is enough to compensate the hard conditions found by dust particles
in  the ICM environment, where the media is dominated by hot gas, and then a relatively
fast sputtering  of dust grains which seems to dominate over the accretion of dust particles. 

First searches for ICM dust were  considered in the seminal work  by \citet{zwi62}, having since then a
long history of detection and refutation.  Two different approaches have been traditionally used:
several groups  compared the possible additional attenuation of light with respect to the field and
subsequent reddening of  objects in the background of clusters. Some of the best limits on this effect
have been obtained by \citet{nol03} and \citet{gut14} (hereafter Paper I) with upper limits on that
additional extinction $E(B-V)$ of the order of a few milimags. Other works follow a different line
trying to measure directly the emission from ICM dust in the far infrared and radio. Due to the small
signals expected, it has been used in general a statistical approach averaging the contribution of many clusters
(Montier \& Giard 2005; Giard et al. 2008; Paper I), although estimations based on single clusters also
exist  \citep{sti02, bai07, kit09}. Some of the results claiming for a detection are controversial, and
there is a broad consensus to interpret them as upper limits on the presence of dust in the ICM.

In Paper I  we used a large  sample of clusters obtained from the Sloan Digital Sky Survey, SDSS \citep{wen12}
complemented at low redshift with the list of Abell clusters with known redshifts, the
catalog of SDSS galaxies with photometric redshift estimations, and the \citet{sch98}
extinction map. In that paper we follow the two methods outlined above; on the one hand  we stacked and
averaged the Schlegel et al.  maps centered in all the clusters of the sample,
applying standard aperture photometry in which the average extinction from the Milky Way was averaged
and subtracted. This
procedure also removed
any other possible emission  in angular scales much larger
than those of the clusters. Alternatively,  we compared the colors of SDSS galaxies with
photometric redshifts in the background of such clusters with other galaxies at similar
redshifts projected through {\it clear} lines of sights. Both methods give compatible results:
a maximum extinction due to the ICM of a few milimags, and upper limits on the dust
fraction in the ICM of  $\sim
10^{-5}$ of the total mass of the cluster. However, as it was noted in Paper I,
there are two major drawbacks that hinder that analysis: firstable,   the relatively
poor angular resolution ($\sim 5$ arcmin) of the extinction map that prevents
the separation of the possible component originated in the ICM from those coming
directly from galaxy members of the clusters, and secondly  the relatively large
uncertainties in the photometric redshifts of SDSS galaxies that make impossible in
the majority of the cases assess the relative  situation  through the line of sight of 
galaxies with respect to the clusters.

In this second paper, we use the same SDSS based catalog of clusters, whilst as tracers
of the infrared
emission of such clusters, we select  the recently released HERSCHEL maps and catalogs
of the HerMES project. These comprise maps centered at three wavelengths, 250, 350 and 500
$\mu$m, and have much  better spatial resolution than the IRAS and DIRBE maps. The
combination of those maps and the corresponding catalogs of  sources allows  to estimate tighter 
upper limits on the ICM fluxes by
subtracting from the total emission of the clusters, the contribution of cataloged sources, or by 
using the
position of these sources to identify specific regions of
the maps projected along the line of sight of a given cluster, free of bright sources (those
included in the catalogs). This, in conjunction with the much better sensitivity of HERSCHEL as
compared to IRAS or DIRBE,  largely compensates the small sky area covered by the HerMES
project. 

The goals of this paper are  the following: i/ to determine the total emission of the
clusters at observed wavelengths of 250, 350 and 500 $\mu$m,  ii/ to estimate 
the fraction of
these signals due to dust in the ICM, iii/ to set upper limits on the extinction produced
by such ICM in the light of background objects, iv/ to study how these magnitudes change with redshift
and/or mass of the cluster, and v/ to determine how the total infrared luminosity and therefore, the star formation
rate (SFR) of the clusters evolve with redshift. To avoid duplication, we refer the 
reader to Paper I for a more detailed description of the samples and
methodology used. Here, we focus on those aspects that have changed
from that paper or are entirely new. The paper  is structured as follows: Section 1 corresponds to this
introduction; Section 2 describes the properties of the samples used, the restrictions applied to get
the final subsamples, and the methodology;  Section 3 presents the main results; Section 4 considers the
evolution with mass and redshift; whilst Section~5 compares our results with those previously found; 
finally  conclusions are presented in Section 6.

\section{Sample and methodology}

The catalog of clusters used is the one by \citet{wen12} as it was done in the  main analysis
of Paper I. Basically, it is a SDSS based catalog that
contains 132,684 clusters with photometric redshifts between 0.05 and 0.8.  The catalog
is above 95\% complete for clusters with $M_{200}>10^{14} M_\odot$ up to redshift 0.42.
For each cluster, the parameters needed for the analysis done in this paper  are the angular position in the sky in right ascension and
declination, the photometric estimation of redshift, and the mass as estimated from the
number and luminosity of member galaxies of the clusters. 
According to the authors of the
catalog, the uncertainties in photometric redshifts are  $<0.03$ for those clusters at
redshifts $<0.42$, and slightly larger for clusters at higher redshifts. 
Although some of the clusters have spectroscopic redshifts, in order to use the whole
catalog in a  consistent way, we decided to use the photometric estimation of redshifts (even in those
cases in which spectroscopic redshifts were available). 

For the estimation of the infrared emission, we have used the HERSCHEL SPIRE \citep{pil10,
gri10} maps \citep{lev10} and catalogs \citep{smi12, ros10, wan14} at 250, 350 and 500 $\mu$m
corresponding to the second data release of the HERSCHEL Multi-tiered Extragalactic Survey
(HerMES) project\footnote{http://hedam.lam.fr} described in \citet{oli12}. The data from the
HERSCHEL mission have angular resolution of 17.6, 23.9 and 35.2 arcsecs at 250, 350 and 500
$\mu$m respectively. For most of the analysis presented in this paper, we will analyze the
three maps independently. The maximum angular resolution of such maps (17.6 arcsecs at 250
$\mu$m) translates into spatial resolutions 15 - 130 Kpc\footnote{Through this paper we use a
cosmological model with $h=0.7$, $\Omega_M=0.3$ and $\Omega_\Lambda=0.7$.} for clusters in the
range of redshifts 0.05 - 0.8. As HerMES covers only a small fraction of the whole sky, the
large majority of the Wen et al.  clusters are not within the area observed by that project.
From the list of 23 maps available in the HerMES Second Data Release (DR2), we have selected
those fields that map the region around at least 10 clusters of the Wen et al.  catalog. After
several tests, we decided to exclude from the analysis also the field L2-COSMOS due to its
relatively high noise as compared to the rest of the maps selected.  The final selection is
listed in Table~1 which shows the name of the fields (column 1) and  their nominal equatorial
coordinates (columns 2 and 3). In each of the maps, we discard for the analysis a small section
that corresponds to edge regions that have not been homogeneously covered; this is done by
selecting  a region  defined by a circle centered approximately in the center of each field,
and with the radius indicated in the fourth column of Table~1. The number of clusters within
each of the fields is listed in column 5. In total, this study uses 37 square degrees of HerMES
maps and 327 clusters (in addition, 18 clusters  projected  in the field of L2-COSMOS and 13 
disseminated through other  HerMES fields were  not used in this work). To estimate the
emission originated in member galaxies of the clusters, we  use  the catalog  obtained by the
HerMES team. The number of detected sources in the 250 $\mu$m channel within a 5 arcmin radius from the center of the
clusters  is presented in the last column of Table~1. The density of
sources projected at clustercentric  distances up to 5 arcmin  is about 27\% higher  than at distances
$5-10$ arcmin, demonstrating that part of the sources correspond to member galaxies
of the clusters.

\begin{figure}
\epsscale{1.}
\plotone{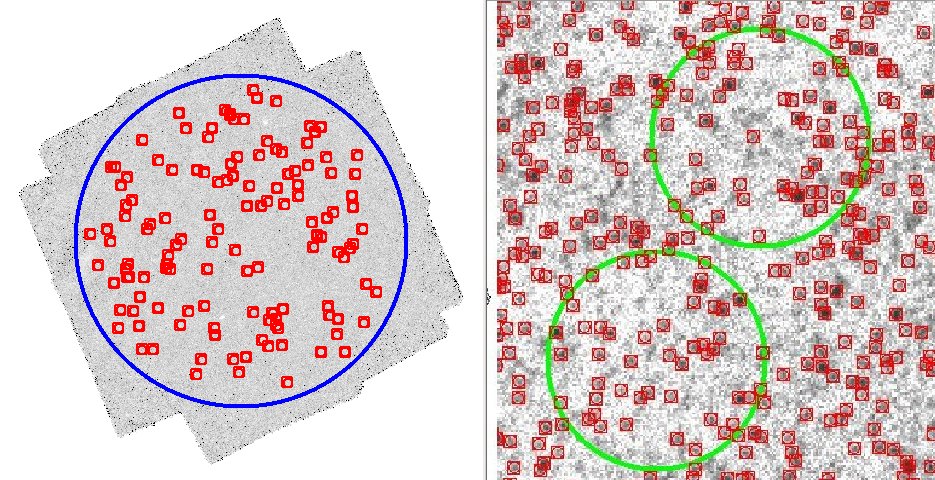}
\caption{Example of one of the HerMES fields (L6-XMM-LSS-SWIRE) used in our analysis.
($Left$:) HerMES map at 250 $\mu$m; the big circle
has radius of 7,800 arcsecs (see Table 1) and encloses  the region of the map selected for our
analysis, whilst the small squares indicate the position of clusters cataloged by
\citet{wen12}. ($Right$:) A section of the  map in
which the two large circles indicate the position of two of the Wen et al. clusters, whilst the
small symbols indicate the position of  sources detected in that map. See the electronic edition of
the journal for a color version 
of this figure.\label{fig0}}
\end{figure}

Fig.~\ref{fig0} presents an example of one of the HerMES maps (L6-XMM-LSS-SWIRE at 250
$\mu$m) used in our analysis. It is shown (left panel) the map and the positions of the 
clusters detected in the Wen et al. catalog; the
panel on the right shows a zoom of the map, with  two circles of 5 arcmin  radius
centered in the position of two clusters, and the  sources detected in the HerMES
catalog. 

\section{Analysis and results}

\subsection{Total emission}

Through the line of sight of a given cluster, the HerMES maps contain the contribution
from diffuse and point source emissions in the Milky Way,  sources (galaxies)
members of the clusters,  field foreground and background galaxies,  any other
hypothetical  component on superclusters scales, and the possible emission from
ICM. To separate  the emission  of the clusters  from the rest of
contributors, we follow a statistical approach in which the flux  in the angular region
projected through each cluster is computed in clustercentric coordinates, and
therefore by 
stacking the emission from many clusters, gradients or spatial irregularities  from the
rest of contributors (that obviously are uncorrelated with the position of the
clusters) are largely filtered out. So, these contaminants  can be considered as a
spatially uniform background, and
subtracted by measuring them in a region far enough from the central position of the
cluster.  The procedure also smooths any possible spatial asymmetries or irregularities
in the emission profile of the clusters, and then in the stacked maps we can ignore any angular dependence and
consider only the radial profiles. 

\begin{figure}
\epsscale{0.8}
\plotone{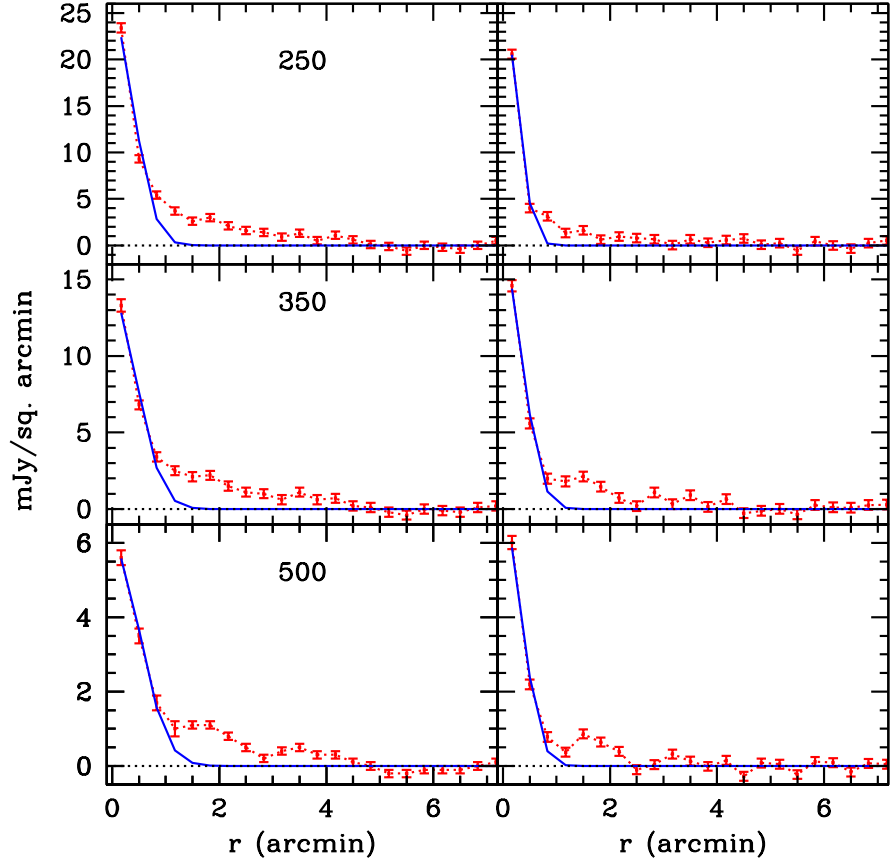}
\caption{$(Left:)$ Mean radial profiles of the cluster emission for the 250 ($top$), 350 ($middle$) and 500
$\mu$m ($bottom$) channels. $(Right:)$ Mean emission from the sources detected in each of these
channels. The solid lines in each panel correspond to
gaussian fits in an inner radius of 1 arcmin. \label{fig2}}
\end{figure}

Figure~\ref{fig2} (left) shows the radial profiles of the average flux per cluster obtained 
by stacking the signal in the 250, 350 and 500 $\mu$m channels for the whole 327
clusters in our sample. The emissions from the clusters are clearly detected in each channel, have a maximum in
the center, and show clear declining radial dependences up to $\sim
3-5$ arcmin from the center, whilst from 5 arcmin onwards  remain nearly flat. The
contribution from contaminants (any other signal apart from the clusters) can be
estimated and subtracted by averaging the signal at distances $5-10$ arcmin from the
center of the clusters\footnote{The figure shows the remaining signals after such
subtraction.}. The signal obtained after such subtraction is our best estimation for the
mean emission of the clusters in the whole sample.  Although considering larger radius
would prevent for removing possible contribution from outskirts of the clusters, in
practice the larger area considered and the comparatively low fluxes would largely
increase the uncertainty in these estimations. 

The spatial profiles of the signals are the result of the convolution of the PSFs of the
HERSCHEL maps with the angular extension of the clusters. Gaussian  fits to the inner part
($<1$ arcmin) of the  radial profiles (solid lines in  Figure~\ref{fig2}) give FWHMs of 56, 65
and 72 arcsecs  for the channels at 250, 350 and 500 $\mu$m respectively. These values are
clearly larger than the corresponding instrumental resolutions, and then indicate that the
clusters are resolved in the maps. The mean redshift of the sample is 0.37, which corresponds to an angular
scale $\sim 5$ Kpc arcsec$^{-1}$, and then radius in the range  $1-5$   arcmin correspond to
linear scales of $0.3-1.5$ Mpc. Table~\ref{tab2} (lines labelled $Total$) shows the integrated
fluxes within such range of radius. The signal is higher in the
high frequency channel, as it is expected for dust with properties similar to the Milky Way. 
Some additional uncertainty on the estimation of the total flux results from the bumps in the
radial profile centered at $\sim 2-4$ arcmin; these produce a considerable increase of the
integrated fluxes from radius 2 to 4 arcmin. These bumps interrupt  the decline of the profile
with radius, and could be the result of the contribution of a set of clusters with particularly
extended profiles, or alternatively the result of fluctuations in the background.

\subsection{Separation of the contribution from galaxies and from ICM}

Obvious upper limits for any of the two components (galaxies and ICM)  are given by the total
emission of the clusters in each channel. However, the sensitivity and spatial resolution of
HerMES allows partially separate and  put tighter constraints on the relative contribution of
both components. To do that, we checked if some of the member  galaxies of the clusters can be
detected in the HerMES maps. Instead of building our own catalog of sources from the maps, we
have used those released by the HerMES group; in particular, those obtained using the 
StarFinder program. Following \citet{wan14} that program is particularly useful deblending and
identifying faint sources in crowded fields as by construction galaxy clusters are. In general,
we do not have information about the redshift of those sources and then it is not possible to
assess if a source projected along the line of sight of a given cluster is a member or not of
such cluster. Therefore, as it has been done for the maps, we used a statistical approach
stacking and averaging the contribution of such sources in clustercentric coordinates. After
subtracting a constant background estimated from the emission within radius $5-10$ arcmin from
the cluster centers, we have obtained the radial profiles presented in the right panels of
Fig.~\ref{fig2}. These profiles indicate that some of the sources detected are members of the
clusters, and by comparing with the profiles of the total emission, that an important fraction
of the emission originated in the cluster comes from these bright discrete sources (galaxies). 

Two estimations of the total flux from a direct integration of the profiles  are presented in 
Table~\ref{tab2} (lines labelled $Galaxies$).
The FWHMs of  gaussian fits to the inner parts $(>1$ arcmin) of the radial profiles are 38, 51 and 49 arcsecs at 250, 350 and 500 $\mu$m
respectively.  These fluxes contain only the contribution from bright galaxies in the clusters
(those bright enough to have been detected and cataloged), so they can be considered as lower
limits to the total contribution from galaxies within the clusters. The radial profiles of the
fluxes due to the cataloged sources are narrower than the corresponding profiles obtained from
the maps. This could be indicative for the presence of some ICM contribution in the outskirts
of the cluster or merely a consequence of segregation of galaxies, i. e. prevalence of low
luminosity, and then uncataloged sources, in the external regions of the clusters.

The contribution from the ICM can be constrained subtracting from the total emission of the
cluster the part that is originated in galaxies. To do that we follow two different estimations
(named A and B respectively). Both provide upper limits on the possible emission of the ICM
because both estimations include the contribution of sources with fluxes below the sensitivity
of the catalog. To subtract the contribution of such sources it would be needed to assumme
specific parameters for the luminosity function of galaxies and how they  evolve with redshift.
This will be considered in a future paper (Guti\'errez et al. in preparation).

{\bf Estimation A:} An upper limit on the contribution by the ICM is obtained by
subtracting from the total emission of the cluster (line $Total$ in Table~\ref{tab2}) the
contribution of galaxies (line $Galaxies$ in Table~\ref{tab2}). The results are also quoted
in that table (line $ICM-A$).

\begin{figure}
\epsscale{0.8}
\plotone{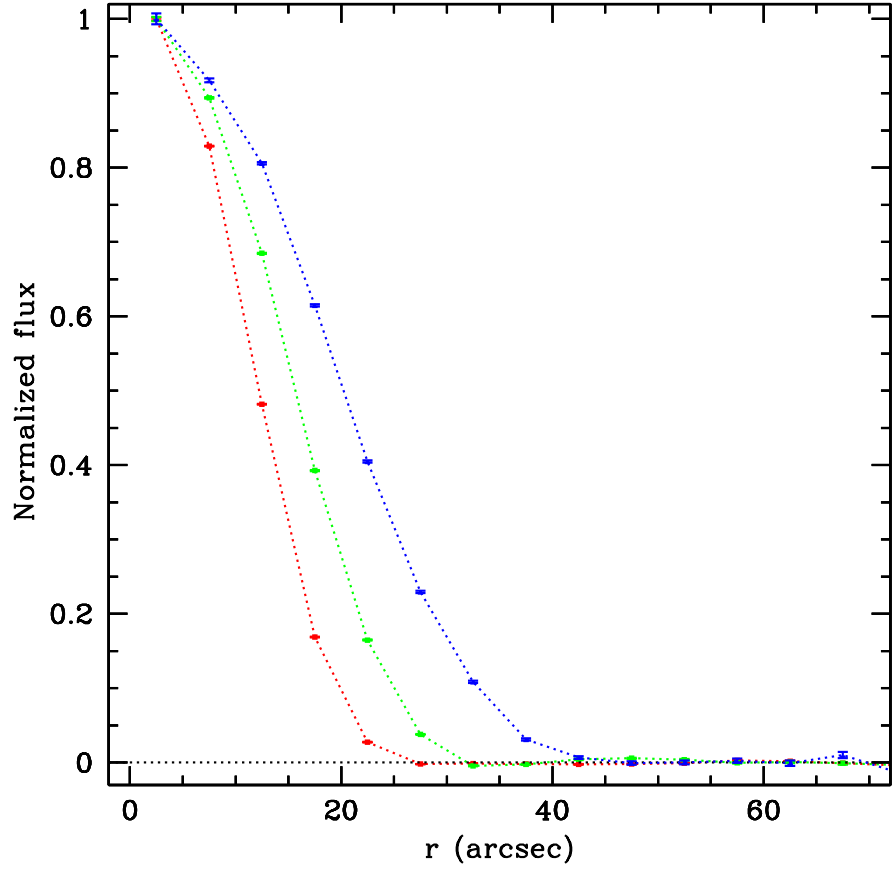}
\caption{Normalized fluxes of pixels as a a function to the projected distance to the nearest
source for the channels at 250  ($red$), 350 ($green$) and 500 $\mu$m ($blue$). \label{fig3}}
\end{figure}

{\bf Estimation B:} Upper limits on ICM fluxes are estimated by stacking and
averaging the fluxes  only in those pixels of the maps that are projected far enough
from any source in the catalog. We estimate which is the threshold in distance by
computing the mean flux of pixels in the maps as a function of the distance to a
given source. Figure~\ref{fig3} presents the normalized radial profiles  of the
fluxes in pixels within distances of 5 arcmin of any cluster, as a function of the
projected distance to the nearest source in the HerMES catalogs. These profiles are
characterized by FWHMs of 24.2, 31.4 and 41.7 arcsecs for the maps at 250, 350 and
500 $\mu$m respectively. These values are significantly larger than the corresponding
instrumental resolutions.  For instance, from the
instrumental resolution and the radial profile shown in Fig.~\ref{fig3} for the 250
$\mu$m channel,  we obtain a FWHM of 16.6 arcsecs as a rough estimation of the {\it
mean} extension of the sources. In the unrealistic case of all the  galaxies being at
the mean redshift of the sample of clusters ($\sim 0.37$), that angular extension
would correspond to a gaussian profile with $\sigma=35$ Kpc. This is a very crude and
biased estimation, because for a given luminosity and radial profile,  galaxies at
lower redshift will contribute more to the flux, and in practice galaxies spread over
a broad distribution of luminosities and profiles which evolve with redshift. Based
on the profiles shown in Fig.~\ref{fig3}, we  defined as regions free of the
contribution of bright sources, and therefore suitable to properly constrain  the
emission of ICM,  those  lying at distances $>30.8$, $>40.1$ and $>53.2$ arcsecs for
the channels at 250, 350 and 500 $\mu$m respectively, from any of the galaxies
detected in the HerMES catalogs\footnote{These correspond to $3\sigma$ values,
approximating the profiles shown in Fig.~~\ref{fig3} to gaussians.}. 

\begin{figure}
\epsscale{0.8}
\plotone{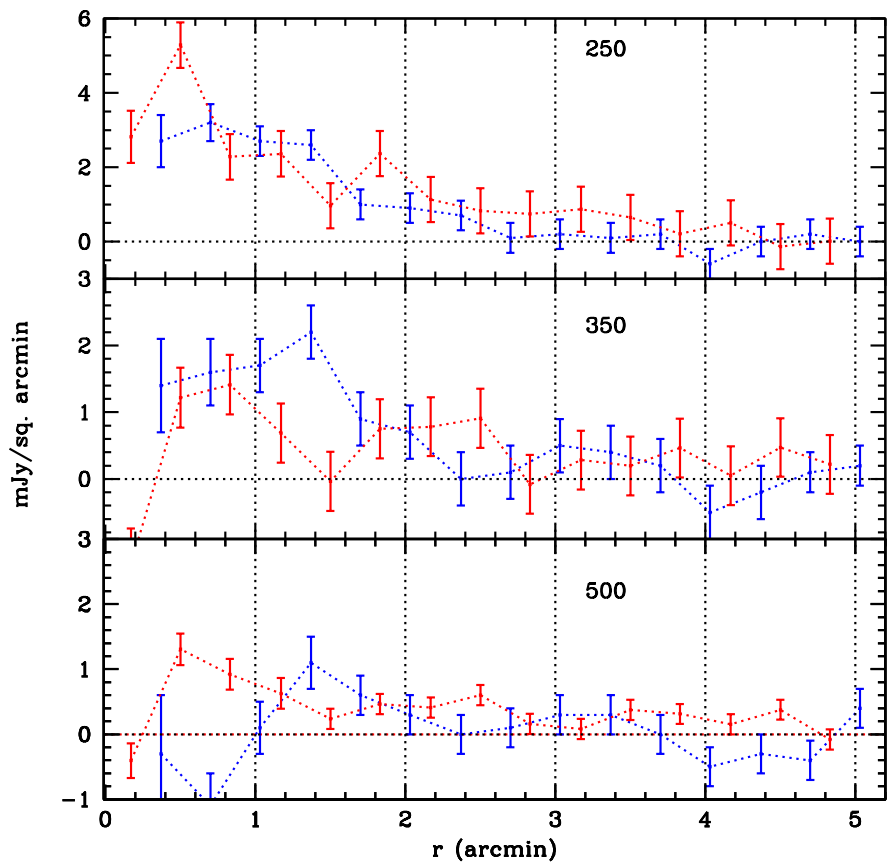}
\caption{Two diferent estimations (A in $red$ and B in $blue$, see main text for details) of the radial profiles for the 250 ($top$), 350 ($middle$)  and 
500 $\mu$m ($bottom$) channels of the emission of the clusters after subtracting the contribution of sources
bright enough to be detected individually. The results for estimation B ($blue$
lines) have been 
slightly
shifted
to the right for visual purposes.
 \label{fig4}}
\end{figure}

Fig.~\ref{fig4} shows the radial profiles  obtained using the A ($red$) and B ($blue$)
estimators for the channels at 250, 350 and 500 $\mu$m.  Both  radial profiles agree quite well
apart from the inner part where the noise is higher (the number of pixels used in the
estimatios is small), and the angular extension of the sources could have some impact on such
estimations. These integrated fluxes  within radius $1-5$ arcmin are presented in
Table~\ref{tab2} under the lines $ICM-A$ and $ICM-B$ respectively. In general, estimations
following $ICM-B$  tend  to be somewhat tighter. These  fluxes represent estimations of
the maximum signal due to ICM, so through the rest of the paper we will follow a 
conservative approach considering the constraints on the ICM obtained with estimator $A$.
From the results obtained  within 5 arcmin using that estimator, the 95\% upper limits
on the ICM  surface emission  are $1.3\times 10^{-2}$, $0.7\times 10^{-2}$, and $0.5\times
10^{-2}$ MJy str$^{-1}$ from the channels at 250, 350 and 500 $\mu$m respectively. 

\subsection{Dust masses in the ICM}

The restriction on the dust mass in the ICM from the above estimation of fluxes requires assumptions of the unknown
emissivity (and then chemical composition) of dust particles and on the temperature of the
media. These estimations of mass differ by roughly an order of magnitude for dust temperatures
within typical  ranges (15-25 K) found in studies of extragalactic galaxies
 \citep{gal12, tab14}.
Adopting a model similar to dust in interstellar regions of the Milky Way (T=20 K and
opacities given by the \citet{dra84} silicate-graphite model) we have obtained
the results of Table~\ref{tab3}. Masses of dust are shown in terms of solar masses and in terms of
the total mass of the clusters as estimated by Wen et al. As these estimations come
from the fluxes obtained using estimator A and therefore contain also the contribution
of faint galaxies, they must be considered as upper
limits of dust mass in the ICM. Using the more restrictive limits found using the channel at 350 $\mu$m,
the mean ICM mass per cluster within 3 arcmin is $< 10^9\, M_\odot$ or $9.5\times
10^{-6}\, M_{cluster}$ (95\% C. L.). The corresponding  limits on the projected
surface densities are 
$1.3\times 10^7\,M_\odot\, arcmin^{-2}$ or $1.2\times 10^{-7}\, M_{cluster}\,
arcmin^{-2}$.

The constraints presented here, imply that the gas to mass ratio in ICM
is several orders of magnitude  smaller than in the Milky Way \citep{dra03}. For instance,
taking  the limits obtained in the inner 5 arcmin, this
defficiency corresponds to $\sim 1.2\times 10^{-3}$ which could be at least partly the
result of the relative short living time of dust in the ICM.

\subsection{Extinction of background objects}

We have estimated upper limits on the  visual extinction produced by ICM following a
 similar approach than in Paper I, i. e.
scaling the extinction ($A_{V,MW}=8.6\times 10^{-5}$ mag) and  luminosity
($L_{BOL,MW}=2.6\times 10^{43}$ erg s$^{-1}$)
of the Milky Way  \citep{dav97}. Considering a
model of dust with temperature 20 K and emissivity 2, we have,

$$L_{BOL}=\int _0 ^\infty L_\nu d\nu=\int _0 ^\infty L_0\nu^2B(\nu, T=20 K)d\nu = 9.44 \times
10^{17} L_0\,  	\Rightarrow L_{0,MW}=2.75\times 10^{18}\, W$$

and for the clusters,

$$L_{\nu _{rest}}=4\pi d_l^2\frac{F_\nu}{1+\bar z}=L_0\nu^2$$

where $L_{\nu_{rest}}=8.89\times 10^{24}$, $7.39\times 10^{24}$ and $3.99\times
10^{24}\, W/Hz$ are the luminosities per unit of rest frequency for the 250, 350 and
500 $\mu$m channels respectively, $\bar d_l$ and $\bar z$ are the mean luminosity distances and redshifts of the
clusters, whilst $F_\nu$ are the limits of the fluxes for ICM emission as
quoted in Table~3.  For the 327 clusters considered in this work, $\bar d_L=2,228$ Mpc and 
$\bar z =0.371$. The upper limits on the extinction  within radius from 1 to 5 arcmin 
 derived from each of the channels are shown in  Table~\ref{tab4}. These values correspond
to maximum values assuming the unrealistic case of negligible contribution from faint sources. In
any case, they are probably the best limits found for the extinction through
clusters. 

\section{Estimating dependences with redshift and mass}

\begin{figure}
\epsscale{0.8}
\plotone{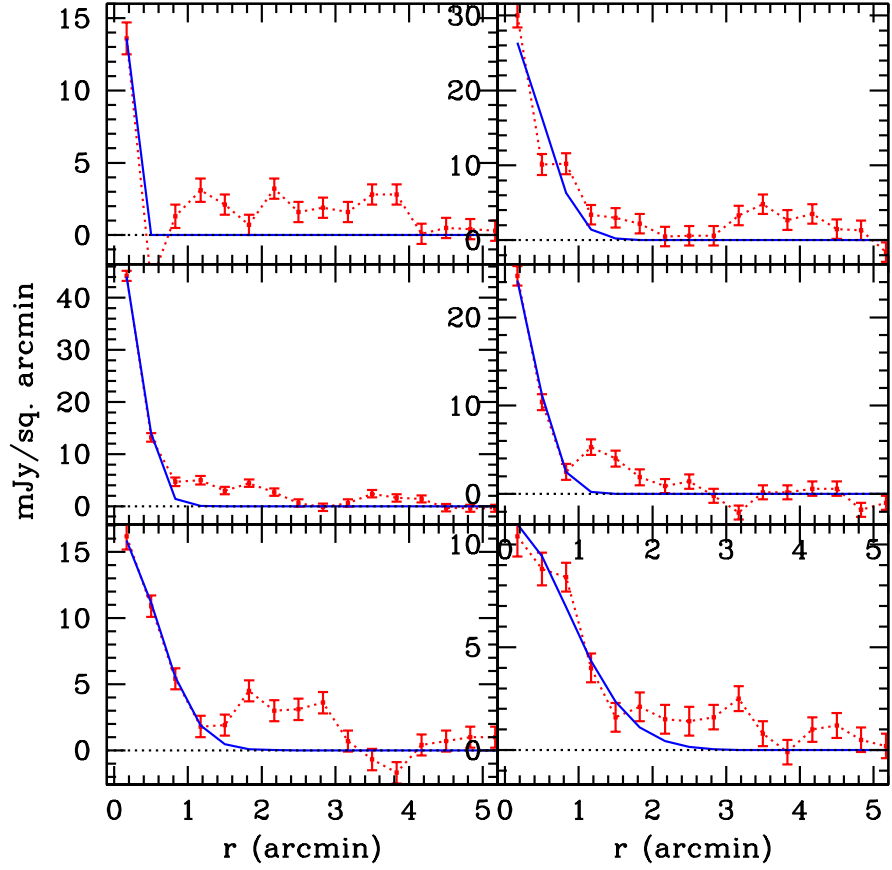}
\caption{Radial profiles of the mean  fluxes in the 250 $\mu$m channel for clusters in two ranges of
masses ($left:$ $<10^{14}$ and $right:$ $>10^{14}\, M_\odot$) and three of redshifts ($top:$ $<0.24$,
$middle:$ $0.24-0.42$ and $bottom:$ $>0.42$). The solid lines correspond to gaussian fits to such
profiles.    \label{fig5}}
\end{figure}

The range in redshift and mass spanned by the sample of clusters used allows to study the
possible evolution of fluxes, masses and extinctions as functions of both parameters,
redshift and mass. That was done by separating the clusters according to three ranges in redshift
$<0.24$, $0.24-0.42$ and $>0.42$, and two in mass ($<10^{14}$ and $>10^{14}M_\odot$) respectively.
We repeat for each of the bins the procedure of stacking, separation of components, and estimation of
physical magnitudes carried out for the whole sample. The radial profiles of the mean
fluxes for each of the division in redshift and mass are presented in Fig.~\ref{fig5}. Although the 
comparatively low number of clusters considered in each bin  respect to the whole
sample obviously increases the noise on the estimation of fluxes, these are clearly detected in each
subgroup of redshift and mass. The gaussian fits to these profiles, shown as solid lines in the figure, 
 are able to
reproduce reasonably well  the inner 1 arcmin in all cases apart from the bin that
corresponds to $<10^{14}M_\odot$ and redshift $< 0.24$. The radial profiles show that clusters tend to be more
extended with redshift. This effect is not present in the estimations of sizes from
the distribution of galaxies in the optical as was done by Wen et al., and could be
indicative of star formation in the outskirts of high redshift clusters by galaxies in the
process of being accreted.

\subsection{Extinction of background objects}

Table~\ref{tab5} presents the upper limits obtained on the visual extinction of background objects  for the
three bins in redshift and two in mass of clusters considered. The columns are the number of clusters in each bin, the mean redshift and
mass, and the 95\% upper limit on the visual extinction within a radius of 5 arcmin obtained
from the 250, 350 and 500 $\mu$m channels respectively. The three channels
independently allow to put constraints at levels below $\sim 1$ milimag for
any of the subdivisions in redshift and mass considered. 

\subsection{Evolution of the infrared luminosity}

Studying the possible evolution of luminosity with redshift and/or mass could give information
about the general processes ruling the accretion of galaxies by  clusters, and the internal
mechanisms that could affect the evolution of galaxies and the star formation processes.
Through this section, we consider the whole luminosity of the cluster ignoring the possible
contribution of the ICM. The analysis below implicitly assummes that  physical conditions and
evolution of dust in  ICM and  galaxies are similar. For this analysis,
we  considered the emission within a 5 arcmin radius. Instead of integrating the profiles of
each bin in the three channels, we have chosen the channel 250 $\mu$m as reference, and 
estimated the fluxes in the other two  channels by scaling the integrated fluxes  at 250 $\mu$m 
 according
to the relative signals in the central bins.  We estimated the corresponding luminosities per unit
of rest framed frequency, $L_{\nu_{rest}}$, considering the luminosity distance $d_l$ at the
mean redshift of each bin, and correcting to a rest framed system, i e. $L_{\nu_{rest}}=4\pi
d_l^2F_\nu/(1+z)$, where $\nu$ is the observed frequency of each of the three channels, and 
$\nu_{rest}=\nu (1+z)$.  The frequencies observed for a given channel correspond to different
rest framed frequencies according to the redshifts of the clusters, therefore, in order to do a
proper comparison of luminosities, it is necessary to estimate the luminosities at a particular
rest framed frequency or to estimate the total far infrared luminosity ($L_{FIR}$) by assuming
a model for the emission of the dust. 

\begin{figure}
\epsscale{0.8}
\plotone{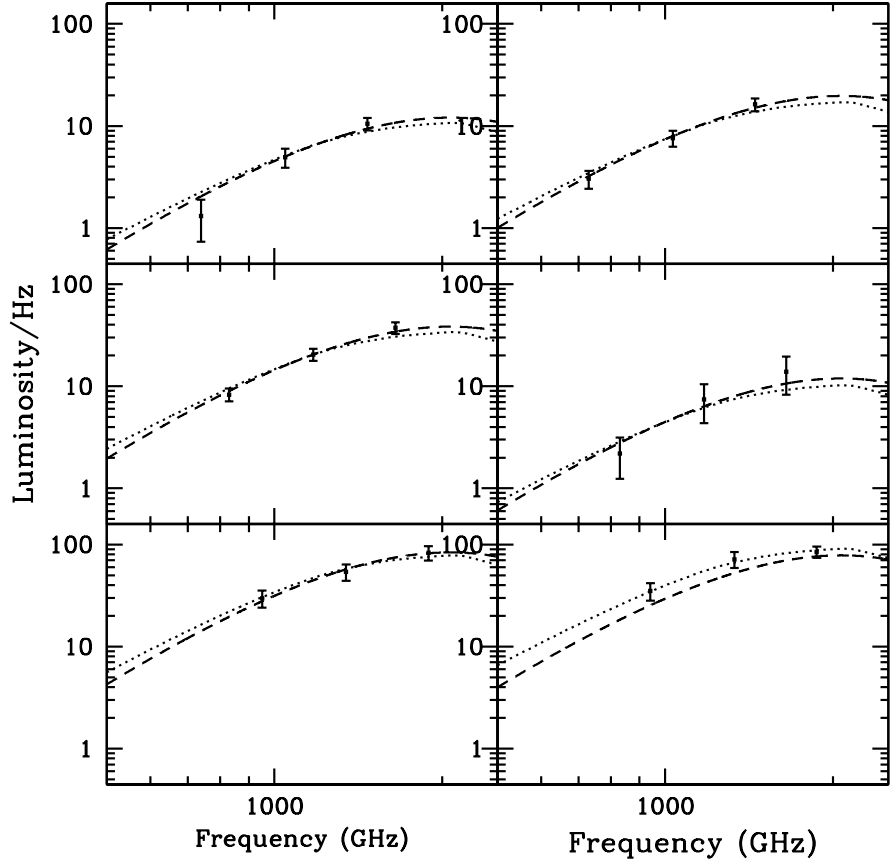}
\caption{Luminosities per unit of frequency  obtained for clusters in the three
ranges of redshift (from top to bottom: $0.05-0.24$, $0.24-0.42$ and $>0.42$), and two of mass (left:
$<10^{14}M_\odot$ and right   $>10^{14}M_\odot$).  The units of the vertical axis
are $erg\, s^{-1} Hz^{-1}/10^{31}$ whilst the horizontal axis corresponds to rest
framed frequencies. Dashed and dotted lines  correspond to the best fits for a model  $\sim \nu^2 B_\nu (T=20
K)$ and  for a \citet{dra07} model respectively (see main text for
details).\label{fig6}}
\end{figure}

A model in which the emission of dust
corresponds to a modified black body with a temperature $T$ and an
emissivity $\beta$, i.~e.  $L=\int L_\nu d\nu=K\int \nu ^\beta B_\nu(T)d\nu$ gives
an adequate description of most of the dust emission at $\lambda \geq 50\,\mu m$. 
Leaving unconstrained  the three parameters ($K,\, \beta$ and $T$), reasonable fits from the estimated
luminosities in the  250, 350 and 500 $\mu$m channels are
obtained for values of $T\sim 20-25\, K$ and $\beta\sim 2$. However, the relatively small
spectral range covered by the three channels, the high noise on the estimation of
luminosities, and the known degeneracy between temperature and emissivity, make 
convenient to reduce the number of free parameters. In the
analysis presented below, we chose $T=20\, K$ and $\beta=2$. Nevertheless, the 
estimations of the relative luminosities do not depend much
on the exact values of these parameters.  Fig.~\ref{fig6} presents the
luminosities per unit of frequency for each of the channels obtained for clusters
in the  ranges of redshift  and mass considered, and  the best fits (dashed lines)
obtained  for
this model. In general, the model reproduces well the data with
predictions within the error bars of the estimated luminosities. The
bolometric luminosities obtained for these models are presented  in Table~\ref{tab6}.
These results show a clear dependence of luminosity with redshift whilst the possible dependence on
mass, if any, is less conspicuous.

The evolution of luminosity  with redshift and mass was modeled according to $L=\Upsilon
M^{\alpha_M} z^{\alpha_z}$, with $\Upsilon$, $\alpha_M$ and $\alpha_z$ parametrizing the
overall normalization, and the possible dependence  with mass and redshift respectively. A chi-square fit gives
$\Upsilon=1.60\times 10^{44}\, erg\, s^{-1}$, $\alpha_M=-0.24$ and $\alpha_z=1.47$.  Given the
number of assumptions and systematic uncertainties associated to this procedure, we have not
tried to quote errors in these parameters. The fit is able to reproduce well the data although
obviously the relatively low luminosity of the intermediate redshift, high mass bin is somewaht
out of the general trend. We have checked that the fluxes measured for this bin in the three channels  are
clearly lower than those in the low mass bin at the same redshift. It would be
interesting to investigate this in the future with more Herschel data, in order
to see if some systematics is present in the estimation of fluxes, or the
results just  show particularities  in the evolution of high mass clusters at
intermediate redshifts. Excluding that bin from the analysis, the fit notably improves with
differences between the data and the fits within $\sim 20\%$. The fit gives 
$\Upsilon=1.73\times 10^{44}\,erg\, s^{-1}$, $\alpha_M=0.36$, and $\alpha_z=1.57$. In any case,
the strong dependence of the
luminosity with redshift is quite robust as it was demonstrated by conducting numerous tests in
which we split the data changing the  number of bins in redshift and mass. We also repeat the
analysis, using the luminosities  estimated from the gaussian fits to each of the radial
profiles\footnote{In this case we exclude from the analysis the  bin that corresponds to the
lower mass and nearest redshifts because its radial profile is too noisy and it was not
possible to obtain a reasonable fit.}. Using those luminosities we obtain $\Upsilon=1.16\times
10^{44}\, erg\, s^{-1}$, $\alpha_M=-0.29$ and $\alpha_z=1.48$. 

More realistic models of dust emission \citep{dra07} show that the modified back body models
underestimate the dust emission at very long wavelengths ($>1000\,\mu m$) and do not describe the
spectral range  shortwards than $\sim 50\, \mu m$, in which emission associated to polycyclic
aromatic hydrocarbon (PAH) material is the dominant process. This is important in order to
estimate $L_{FIR}$, which is used as an estimator of the $SFR$ \citep{ken98} (see below). The
relative emission at short wavelengths depends on the amount of PAHs and on the properties of the
radiation field. As representative of these models, we fit the luminosities in each of the 3 x 2
samples in redshift and mass respectively to a Milky Way like model with a fraction of PAHs of
4.6\%, and exposed to a single radiation intensity as computed by \citet{dra07} (see that paper
for full details). Fitting the luminosities from the integrated  fluxes within 2 arcmin in the
three channels, we obtained the fits  presented in Fig.~\ref{fig6} (solid lines). The figure
shows that these models reasonably fit the data and  predict emission similar to the modified
black body model in the range $\sim 50-500\,\mu m$. We are aware of the uncertainties in the estimation of $L_{FIR}$  due to the
contribution of dust at wavelengths shorter than $\sim 50\, \mu m$ that are not sampled by the
HERSCHEL data, however this uncertainty is mostly systematic and likely affects in a similar way to the six subsamples.
Repeating the above analysis with the
luminosities estimated from the \citet{dra07} model, we obtain $\Upsilon=1.23\times 10^{44}\,
erg\,
s^{-1}$, $\alpha_M=-0.16$, $\alpha_z=1.63$, and $\Upsilon=1.73\times 10^{44}\, erg\, s^{-1}$,
$\alpha_M=-0.33$, $\alpha_z=1.53$ for the luminosities obtained integrating directly the radial
profiles within a 5 arcmin radius, and from the gaussian fits respectively. To study separately the evolution with redshift for the low and high mass samples, we
fit functions $L\sim z^{\alpha_z}$ (i. e. $\alpha_M=0$) for the low and high  mass subsamples and obtained
$\alpha_z=1.93$ and $1.35$ respectively.

To check how the parameters $\alpha_M$ and $\alpha_z$ depend on the model used to parametrize the cluster luminosities,
we also simply fit a straight line to the luminosities derived for each of the
three channels and subsamples of redshift and mass, and take the interpolate value
at a given frequency. Using the luminosities at $\nu_{rest}=1250\, GHz$, we obtain
$\alpha_M=-0.1$ and $\alpha_z=1.5$. 

The dependence of luminosity with redshift and mass translates directly into SFR; using
\citet{ken98} calibration, i.~e. $SFR\,(M_\odot\, yr^{-1})=4.5\times 10^{-44}\, erg\,
s^{-1}L_{FIR}$ (see \citet{ken12} for a discussion and comparison between different SFR estimators), our results show a strong declining of $SFR$ for clusters in both ranges of
masses ($<10^{14}$ and $>10^{14}$ M$_\odot$) by nearly an order of magnitude in a period of
$\sim 6$ Gyr. This estimation  ignores the contribution to luminosity from AGNs; however we implicitly
assume a coevolution of black holes and $SFR$ in, at least, the last 10 Gyr,see e. g. 
\citet{sha09}  and the evidence summarized in \citet{hec14}. So,
ignoring AGNs contribution, the absolute calibration of SFR could be affected but in the 
same way at any redshift. The little dependence of $L_{FIR}$ with mass indicates a specific star formation
rate, $SFR/M $ anticorrelated with mass (excluding the intermediate redshift high mass bin, we obtain
$SFR/M\sim M^{-0.64}$). This anticorrelation support, as many other authors have shown in the past
(e. g. \citep{bal98} that the declining of
$SFR$ in clusters with time is not merely a consequence of accretion of galaxies with less
$SFR$ at a given cosmic time, but instead it is also a consequence of internal  processes in
the clusters. 

\section{Comparison with previous studies} 

A similar approach to our method B was used by \citet{kit09} and applied to ISO data at 70 and 160
$\mu$m in the Coma cluster. These authors obtained $A_V<0.011$ mag within the central 100 Kpc of the
cluster. It is not possible to conduct a direct comparison between their results and our findings
(Tables~4 and 5) due to the use of different samples, and the slightly different way to estimate the
extinctions. However, both results point out to a very small extinction within the central parts of
the clusters, being our upper limits an order of magnitude tighter. We also improve  by roughly and
order of magnitude the constraints obtained by \citet{mul08}. Our results are also compatible and
improve the limits ($E(B-V) <8\times 10{-3}$ mags on 1 Mpc scales) found by the \citet{bov08}. The
differential extinction found in the analysis by \citet{che07} from the  differential reddening of
galaxies in the background of clusters found in the analysis by \citet{che07} (see their Fig. 3)
seems to occur at clustercentric distances $>2R_{200}$. As was noted in Paper I, our results are
difficult to fit with those by \citet{gee10}  who analyzed a sample of low redshift groups and
clusters, and found extinctions $\sim 10^{-2}$ mag on scales $\sim 10$ Mpc. At the mean redshift of
our sample, 5 arcmin subtends $\sim 1.5$ Mpc, and assumming a linear decrease of dust attenuation
with clustercentric distance, the results by such authors (see their Fig. 4) would correspond to
$A_v\sim 0.002$ mag. This is still about one order of magnitude smaller that the typical values quoted
in our Table 4. Although Fig. 4 by McGee et al.  seems indicate a shallower behaviour of the differential
extinction in the very inner parts of the cluster, it is difficult a strong assesment on
this due to the low  angular/spatial resolution  of such results. 

Comparing with the results presented in Paper I, the use of three frequencies, and the best sensitivity and
spatial resolution of the HerMES maps with respect to
IRAS, largely compensates the smaller spatial coverage (and then the numbers of clusters
used) in the present work. Considering as representative of the results obtained in Paper
I, the 95\% upper limit of $E(g-i)=3.6$ milimags found
within the central 1 Mpc, the constraints presented here (Table~\ref{tab4}) improve that by roughly an order of magnitude.

Our results qualitative agree with many previous works that have estimated the infrared
luminosity up to redshift $\sim 1$. They show a strong increase of luminosity, and therefore
$SFR$,
with redshift, in the field, in groups and in clusters, being unclear whether or not the evolution
is similar in any of these environments. Although it is difficult to establish rigurous
comparison due to the different methods, samples, and redshifts covered, the evolution found in
the present work is stronger than the dependence $\sim (1+z)^{3.9}$ found by \citet{lef05} in
field galaxies up to $\sim 1$ found by Spitzer \citet{wer04} in the Chandra Deep Field South.
Other authors, \citep{dye10} have found in field galaxies a stronger dependence in luminosity
with redshift $\sim (1+z)^7$ that  agrees better with our results.

We qualitatively agree also with the findings by \citet{guo14} and \citet{pop12} who found increase
in luminosity with redshift in groups and clusters with masses in the ranges
$(10^{12}-10^{14})h^{-1}M_\odot$ and $(10^{13}-10^{15})M_\odot$ respectively, and with the strong
evolution in luminosity  found also by \citet{bai09} comparing local and high redshift ($\sim 0.8$)
clusters. \citet{cle14}  by analyzing Herschel-SPIRE maps and the Planck Early Release Compact
sources catalog \citep{pla11} identified and estimated the luminosities and $SFRs$ of four
candidates to be clusters of galaxies  at redshifts $\sim 1-2$. Although these sources correspond
likely  to very extreme cases that are not representative of the average values found in the
analysis of Wen et al clusters, their own existence points towards an increase in luminosity,
and therefore  in $SFR$, very strong at least up  to redshift $\sim 2$.

\section{Conclusions}

Constraints on the ICM emission at 250, 350 and 500 $\mu$m have been
obtained using  a sample of 327 clusters of galaxies spanning  ranges of 0.05-0.7 in
redshift and  $>4\times 10^{13}M_\odot$ in mass, and maps and catalogs of 
sources  at 250, 350 and 500 $\mu$m from the HERSCHEL HerMES project.  The main
results are:

\begin{enumerate}

\item  Subtracting from the whole emission of the cluster the contribution of identified
 sources, it has been estimated 95\% upper limits  on the ICM  emission  per cluster of 86.6,
48.2 and 30.9 mJy in the maps at 250, 350 and 500 $\mu$m respectively within a 5 arcmin
radius centered in each  cluster. These correspond to  surface emission of $1.3\times
10^{-2}$, $0.7\times 10^{-2}$, and $0.5\times 10^{-2}$ MJy str$^{-1}$  respectively.

\item Assuming for the intracluster dust typical values found within the Milky Way
(a temperature of 20 K, and an emissivity $\beta=2$), we have obtained strong
upper limits on the visual extinction of background objects due to the ICM. From the fluxes at 250, 350 and 500 $\mu$m these restrictions are 
respectively $A_V(95 \%\,C.L.) <$ 0.4, 0.4 and 0.6 milimags within projected
distance of 5 arcmin from the cluster centers. Separating the clusters in 3 x 2 bins in redshift and mass
respectively,  the  constraints on
the extinctions are of the order of 1 milimag in any of the ranges in redshift and mass considered. 

\item Tight upper limits are also obtained on the  dust mass 
in the ICM; for instance using the data at 350 $\mu$m we set a 95\% upper
limit of $<10^9 M_\odot$ within a projected 5 arcmin radius circle  centered in the clusters.
This corresponds to a fraction of the total mass of the clusters of $9.5\times 10^{-6}$.

\item From the results obtained in the three channels, we have estimated $L_{FIR}$
and its dependence of the 
far infrared luminosity with redshift and mass. Although there are many assumptions
and the estimations are slightly model dependent, the overall results show a clear and strong dependence with redshift
$\sim z^{1.5}$, whilst the dependence with mass is not so obvious indicating an anticorrelation of $L_{FIR}/M$ with
the mass of the cluster. Using the linear relation between $L_{FIR}$ and $SFR$, the results indicate a
 declining by roughly an order of magnitude in $SFR$ in a period of $\sim 6$ Gyr. The specific
 $SFR$ is
 clearly anticorrelated with mass; this indicates  that internal 
mechanisms play a role on this quenching  of $SFR$.

\end{enumerate}
   
\acknowledgments

This research has made use of data from HerMES project
(http://HerMES.sussex.ac.uk/). HerMES is a HERSCHEL Key Programe utilizing
Guaranteed Time from the SPIRE instrument team, ESAC scientists and a mission
scientist. The HerMES data was accessed through the HERSCHEL Database in Marseille
(HeDaM - http://hedam.lam.fr) operated by CeSAM and hosted by the Laboratoire
d'Astrophysique de Marseille.

\clearpage                 

\begin{table}
\begin{center}
\caption{HerMES fields used in this work \label{tbl-1}}
\begin{tabular}{lrrccc}
\tableline\tableline
Field & RA (hh:mm:ss) & Dec. ($^{o}$ $^{'}$ $^{''})$ & Radius (arcsec) & Clusters & Sources
\\
\tableline
L5-Lockman-SWIRE    & 10 46 47.61 & +58 04 38.0 & 6000 &  72 &2467\\
L5-Bootes-HerMES    & 14 32 37.03 & +34 10 17.6 & 5400 &  80 & 3827\\
L5-ELAIS-N1-HerMES  & 16 10 11.02 & +54 19 30.3 & 3000 &  16 & 826\\
L6-XMM-LSS-SWIRE    &  2 20 35.99 &  -4 31 43.4 & 7800 & 129 & 5177\\
L6-FLS              & 17 16 12.74 & +59 23 01.7 & 4200 &  30 &1272\\
\tableline
\end{tabular}
\end{center}
\end{table}

\begin{table}
\begin{center}
\caption{Integrated fluxes of the average emission of 327 clusters.\label{tab2}}
\begin{tabular}{ccrrrr}
\tableline\tableline
$Channel$ & $Radius$ & $Total$ & $Galaxies$ & $ICM-A$ & $ICM-B$ \\
 & $(arcmin)$ & \multicolumn{4}{c}{$(mJy)$} \\
\tableline
250  & 1 & $  27.3  \pm  0.9  $ & $   16.8  \pm  1.0 $ & $ 10.5 \pm  1.3  $ & $  9.0 \pm  0.9 $ \\
  & 2 & $  55.8  \pm  2.5  $ & $   27.7  \pm  2.7 $ & $ 28.1 \pm  3.7  $ & $ 21.9 \pm  2.4 $ \\
  & 3 & $  81.8  \pm  4.5  $ & $   40.0  \pm  5.0 $ & $ 41.8 \pm  6.7  $ & $ 26.8 \pm  4.3 $ \\
  & 4 & $ 101.6  \pm  6.9  $ & $   47.4  \pm  7.7 $ & $ 54.2 \pm 10.3  $ & $ 24.4 \pm  6.5 $ \\
  & 5 & $ 118.2  \pm  9.7  $ & $   60.4  \pm 10.7 $ & $ 57.8 \pm 14.4  $ & $ 26.7 \pm  9.0 $ \\
\\
350  & 1 & $  17.7  \pm  0.6  $ & $   14.4  \pm 0.7  $ & $  3.3 \pm  0.9  $ & $  5.1 \pm  0.9 $ \\
  & 2 & $  38.7  \pm  1.6  $ & $   31.1  \pm 1.9  $ & $  7.6 \pm  2.5  $ & $ 16.0 \pm  2.3 $ \\
  & 3 & $  57.0  \pm  2.9  $ & $   41.8  \pm 3.5  $ & $ 15.2 \pm  4.5  $ & $ 19.3 \pm  4.0 $ \\
  & 4 & $  73.7  \pm  4.5  $ & $   51.6  \pm 5.4  $ & $ 22.1 \pm  7.0  $ & $ 19.5 \pm  6.0 $ \\
  & 5 & $  82.3  \pm  6.2  $ & $   53.5  \pm 7.5  $ & $ 28.8 \pm  9.7  $ & $ 20.0 \pm  8.3 $ \\ 
\\
500  & 1 & $   8.6  \pm  0.3  $ & $    5.8  \pm 0.3  $ & $  2.8 \pm  0.4  $ & $ -1.0 \pm  1.0 $ \\
  & 2 & $  18.7  \pm  0.9  $ & $   11.8  \pm 0.7  $ & $  6.9 \pm  1.1  $ & $  4.6 \pm  2.1 $ \\
  & 3 & $  26.0  \pm  1.6  $ & $   13.3  \pm 1.3  $ & $ 12.7 \pm  2.1  $ & $  6.4 \pm  3.4 $ \\
  & 4 & $  34.6  \pm  2.4  $ & $   16.2  \pm 2.0  $ & $ 18.4 \pm  3.1  $ & $  4.3 \pm  5.0 $ \\
  & 5 & $  38.0  \pm  3.3  $ & $   15.7  \pm 2.8  $ & $ 22.3 \pm  4.3  $ & $  1.2 \pm  6.7 $ \\
\end{tabular}
\end{center}
\end{table}

\begin{table}
\begin{center}
\caption{Upper limits on the dust mass in the ICM.\label{tab3}}
\begin{tabular}{cccc}
\tableline\tableline 
$Channel$ & $Radius$ & \multicolumn{2}{c}{$M_{dust}$} \\
 & $(arcmin)$ & $10^7 M_\odot$ & $10^{-7}M_{cluster}$ \\
\tableline
  250  &     1 & $  13.8 \pm  1.7  $ & $  12.5 \pm    1.6  $ \\ 
       &     2 & $  36.9 \pm 4.9   $ & $  33.5 \pm    4.4  $ \\ 
       &     3 & $  54.8 \pm 8.8   $ & $  49.9 \pm    8.0  $ \\ 
       &     4 & $  71.1 \pm 13.5  $ & $  64.7 \pm   12.3  $ \\ 
       &     5 & $  75.8 \pm 18.9  $ & $  69.0 \pm   17.2  $ \\ 
  350  &     1 & $   7.2 \pm  2.0  $ & $   6.5 \pm    1.8  $ \\ 
       &     2 & $  16.5 \pm  5.4  $ & $  15.0 \pm    4.9  $ \\ 
       &     3 & $  33.0 \pm  9.8  $ & $  30.0 \pm    8.9  $ \\ 
       &     4 & $  48.0 \pm 15.2  $ & $  43.7 \pm   13.8  $ \\ 
       &     5 & $  62.5 \pm 21.1  $ & $  56.9 \pm   19.2  $ \\ 
  500  &     1 & $  14.2 \pm  2.0  $ & $  12.9 \pm    1.8  $ \\ 
       &     2 & $  34.9 \pm  5.6  $ & $  31.8 \pm    5.1  $ \\ 
       &     3 & $  64.3 \pm 10.6  $ & $  58.5 \pm    9.7  $ \\ 
       &     4 & $  93.2 \pm 15.7  $ & $  84.7 \pm   14.3  $ \\ 
       &     5 & $ 112.9 \pm 21.8  $ & $ 102.7 \pm   19.8  $ \\ 
\tableline
\end{tabular}
\end{center}
\end{table}

\begin{table}
\begin{center}
\caption{Maximum visual extinctions ($A_V$) for objects in the background of clusters
derived from the constraints in the ICM  flux at 250, 350 and 500 $\mu$m. \label{tab4}}
\begin{tabular}{lccccc}
\tableline\tableline
Channel & \multicolumn{5}{c}{$A_{V}$ (milimag)} \\
\hline
 & $1^{'}$ & $2^{'}$ & $3^{'}$ & $4^{'}$ & $5^{'}$  \\
\tableline

250 &  1.7 &  1.1 & 0.8 & 0.6 &  0.4 \\
350 &  1.1 &  0.7 & 0.6 & 0.5 &  0.4 \\
500 &  0.7 &  0.9 & 0.7 & 0.6 &  0.6 \\
\tableline
\end{tabular}
\end{center}
\end{table}

\begin{table}
\begin{center}
\caption{Extinctions of background objects as a function of redshift and mass of the
cluster. The three last columns correspond to the 95\% upper limits obtained for $A_V$ within a 5 arcmin
radius from the center of the clusters, using the results from the 250, 350 and 500 $\mu$m channels
respectively. \label{tab5}}
\begin{tabular}{cccccc}
\tableline\tableline
Clusters & Redshift & (Mass/$10^{14} M_\odot$) & \multicolumn{3}{c}{$A_{V}$ (milimag)}   \\
\tableline
   41  & 0.189 & 0.74    &    0.1 & 0.3 & 0.6 \\
   35  & 0.174 & 1.59    &    0.2 & 0.3 & 0.2 \\
   73  & 0.335 & 0.74    &    0.5 & 0.6 & 0.9 \\
   52  & 0.334 & 1.55    &    0.2 & 0.1 & 0.1 \\
   72  & 0.528 & 0.75    &    1.0 & 0.4 & 1.1 \\
   54  & 0.513 & 1.56    &    0.4 & 0.5 & 0.7 \\
\tableline
\end{tabular}
\end{center}
\end{table}

\begin{table}
\begin{center}
\caption{Luminosities  in three ranges of redshift ($<0.24$; $0.24-0.42$ and $>0.42$) and two
of mass ($<10^{14}$ and $>10^{14}M_\odot$).\label{tab6}}
\begin{tabular}{ccccc}
\tableline\tableline
Range in mass & & \multicolumn{3}{c}{Range in redshift}  \\
\hline
	  & & 0.06-0.24 & 0.24-0.42 & 0.42-0.71 \\
\hline
$<1$ & ${\it  1.}$ & 0.185 & 0.333 & 0.527 \\
     & ${\it  2.}$ & 0.74  & 0.74 & 0.75  \\
     & ${\it  3.}$ & 41    & 73   & 72   \\
     & ${\it  4.}$ & 2.9  & 9.2  & 20.1  \\	
\hline
$>1$ & ${\it  1.}$ & 0.173 & 0.338 & 0.517 \\
     & ${\it  2.}$ & 1.59  & 1.55  & 1.56  \\ 
     & ${\it  3.}$ & 35    & 52    &  54   \\ 
     & ${\it  4.}$ & 4.7  & 2.9 & 18.8 \\	 
\tableline 
\end{tabular}
\tablecomments{\rm Lines denoted by {\it 1.,\, 2.,\, 3.}, and {\it 4.} present the average redshift, average mass,
the number of clusters and the  luminosity in each  of the bins considered. The masses
are given in $10^{14}M_\odot$ and the luminosities in $10^{44} erg\, s^{-1}$.}
\end{center}
\end{table}

\end{document}